\documentclass{elsart}
\usepackage{epsfig}
\usepackage{amsmath}
\usepackage{amsbsy}
\usepackage{amsfonts}
\begin{document}
\begin{frontmatter}
\title{Numerical studies of \\ stabilized Townes solitons}

\author{Gaspar D. Montesinos and V\'{\i}ctor M. P\'erez-Garc\'{\i}a}

\address{Departamento de Matem\'aticas, E.T.S.I. Industriales, \\
  Universidad de Castilla-La Mancha,
Ciudad Real, 13071 Spain.}

\maketitle

\begin{abstract}
We study numerically stabilized solutions of the two-dimensional
Schr\"{o}dinger equation with a cubic nonlinearity. We discuss in
detail the numerical scheme used and
 explain why choosing the right numerical strategy is very important
  to avoid misleading results. We
show that stabilized solutions are Townes solitons, a fact which
had only been conjectured previously. Also we make a systematic
study of the parameter regions in which these structures exist.

\end{abstract}


\runtitle{Numerical studies of stabilized Townes solitons}

\end{frontmatter}

\section{Introduction}

In this paper we study systems ruled by the two-dimensional
nonlinear Schr\"o\-din\-ger equation (NLSE) \cite{Sulem,VVz} with
a cubic time-dependent nonlinearity. More precisely we look for
solutions of the Cauchy problem
\begin{subequations}
\begin{eqnarray}
  \label{eq:gpe3}
  i \frac{\partial u}{\partial t} & = &
  \left[-\frac{1}{2}\triangle + g(t)|u|^2\right]u \\
  u(x,0) & = & u_0(x) \in H^1(\mathbb{R}^2)
\end{eqnarray}
\end{subequations}
where $u(x,t): \mathbb{R}^2\times \mathbb{R}^+\rightarrow
\mathbb{C}$, $\triangle =
\partial^2/\partial x_1^2 + \partial^2/\partial x_2^2$ and
$g(t)$ is a real function (the nonlinear coefficient) so that if
$g<0$ the nonlinearity is attractive whereas for $g>0$ the
nonlinearity is repulsive.

When $g$ is a real constant Eq. \eqref{eq:gpe3} is the cubic NLSE,
which is one of the most important models of mathematical physics,
with applications in very different fields \cite{Sulem,VVz}.

It is well known that for $g<0$ if $N\equiv\|u_0\|_2^2$, is above
a threshold value $N_c$, solutions of Eq. \eqref{eq:gpe3} can
self-focus and become singular in a finite time. This phenomenon
is called \emph{wave collapse} or \emph{blowup of the wave
amplitude}. More precisely, there is no blowup when $N<N_c$ but
for any $\epsilon > 0$, there exist solutions with
$N=N_c+\epsilon$ for which there is blowup
\cite{Fibich,Weinstein}.

Eq. \eqref{eq:gpe3} admits stationary solutions of the form
$u(x,t)=e^{i\mu t} \Phi_{\mu}(x)$, where $\Phi_{\mu} (x)$ verifies
\begin{equation}\label{soliton}
\triangle \Phi_{\mu} -2\mu \Phi_{\mu}- 2 g |\Phi_{\mu}|^2
\Phi_{\mu}=0.
\end{equation}
As it is precisely stated in \cite{Sulem}, when $g$ is negative,
for each positive $\mu$ there exists only one solution of Eq.
\eqref{soliton} which is real, positive and radially symmetric and
for which $\int |\Phi_{\mu}|^2 d^2x$ has the minimum value between
all of the possible solutions of Eq. \eqref{soliton}. Moreover,
the positivity of $\mu$ ensures that this solution decays
exponentially at infinity. This solution is called the
\emph{ground state} or \emph{Townes soliton}. We will denote this
solution as $R_{\mu}(r)$ which satisfies
\begin{eqnarray}\label{solitonR}
\triangle R_{\mu} -2\mu R_{\mu}- 2 g R_{\mu}^3=0 \\
\label{boundary} \lim_{r\rightarrow{\infty}} R_{\mu}(r)=0,\
R_{\mu}'(0)=0.
\end{eqnarray}
Once the value of $\mu$ is fixed, the power and width of $R_{\mu}$
are given by $N_{\mu}=\int |R_{\mu}|^2 d^2 x$ and $W_{\mu}=(\int
|R_{\mu}|^2 r^2 d^2 x)^{1/2}$ respectively. By applying scaling
transformations to $R_{\mu}(r)$ it is possible to build a family
of Townes solitons having the same shape and norm but different
widths
\begin{equation}\label{scaling}
R_{\mu} (r)=\mu^{1/2} R_1 (\mu^{1/2} r),\ W_{\mu}=W_1/\mu^{1/2}.
\end{equation}
The equation which verifies the normalized soliton $R_{\mu, N}
(r)=R_{\mu} (r)/N_{\mu}^{1/2}$ is
\begin{equation}\label{solitonN}
\triangle R_{\mu, N} -2\mu R_{\mu, N}- 2 g N_{\mu} R_{\mu, N}^3=0.
\end{equation}
From the theory of nonlinear Schr\"{o}dinger equations it is known
that the Townes soliton has exactly the critical power for blowup
$N_c$, therefore, it separates in some sense the region of
collapsing and expanding solutions. Moreover, the Townes soliton
is \emph{unstable}, i.e. small perturbations of this solution lead
to either expansion of the initial data or blowup in finite time.

The case where $g$ is not constant but a continuous periodic
function of $t$ has arisen recently in different fields of
applications of Eq. (\ref{eq:gpe3}). The intuitive idea is that
(oscillating) bound states could be obtained by combining cycles
of positive and negative $g$ values so that after an expansion and
contraction regime the solution could come back to the initial
state. In this way some sort of pulsating trapped solution, i.e.,
a \emph{breather}, could be generated.

This idea was first proposed in the field of nonlinear Optics
\cite{Berge}. In that context, two-dimensional solitary waves have
been proposed for optical systems \cite{Berge,Malomed,Malomed2}.
There, a spatial modulation of the Kerr coefficient (the
nonlinearity) of the optical material is used to prevent collapse
so that the beam becomes collapsing and expanding in alternating
regions and is stabilized in average. The same idea has been used
in the field of matter waves in Refs. \cite{pisaUeda,pisaBoris}.
In Ref. \cite{Gaspar} some general results are provided for
generic nonlinearities. Finally in Ref. \cite{Humberto} stabilized
vector solitons have been proposed and studied.

The aim of this paper is to study in more detail the stabilization
mechanism. This paper is organized as follows: First, in  Sec.
\ref{method} we present the numerical method used to integrate the
equations, which involves a Fourier pseudospectral scheme, a
split--step scheme and the use of an absorbing potential which
allows to get rid of the radiation. In Sec. \ref{stabilization} we
discuss the phenomenon of stabilization of solutions of the NLSE
and show quantitatively that the structure that remains stabilized
is a Townes soliton, thus confirming the conjecture raised in Ref.
\cite{Gaspar}. We also study the robustness of the stabilized
soliton under parameter variations. Finally, in Sec.
\ref{Conclusions} we summarize our conclusions.

\section{Numerical methods}\label{method}

\subsection{Fourier pseudoespectral scheme}

In this paper we study Eq. \eqref{eq:gpe3} numerically. To this
end we have developed a Fourier pseudospectral scheme for the
discretization of the spatial derivatives combined with a
split-step scheme to compute the time evolution. Split-step
schemes are based on the observation that many problems may be
decomposed into exactly solvable parts and on the fact that the
full problem may be approximated as a composition of the
individual problems. For instance, the solution of partial
differential equations of the type $\partial_t u(x,t) =
N\left(t,x,u,\partial_x, ....\right) u = (A+B) u$ can be
approximated from the exact solutions of the problems $\partial_t
u  =  A u,$ and $\partial_t u = B u$.

To approximate properly the solutions of Eq. \eqref{eq:gpe3} we
include an absorbing potential in the border of the simulation
region to extract the outgoing radiation and provide some sort of
local transparent boundary conditions. This potential may be
included in Eq. \eqref{eq:gpe3} as a complex term
\cite{Tran,Fevens} so that the new equation we will discretize is
\begin{equation}
  \label{eq:gpeabsorbe}
  i \frac{\partial U}{\partial t} =
  \left[-\frac{1}{2}\triangle + g(t)|U|^2-iV_a\right]U,
\end{equation}
where $V_a(x)$ is a positive real function which must be chosen to
maximize the absorption of the radiation while minimizing the
effect on trapped structures (i.e. to mimic the behavior of true
transparent boundary conditions). Moreover, $U(x,t) \simeq u(x,t)$
will satisfy specific boundary conditions on the border of the
simulation region $\Omega_h$. As $\Omega_h \rightarrow
\mathbb{R}^2$ and $V_a \rightarrow 0$ we expect $U \rightarrow u$.

Let us decompose the evolution operator in Eq.
\eqref{eq:gpeabsorbe} by taking \cite{Taha,Xiaoyan}
\begin{subequations}
\begin{eqnarray}
A & = & \frac{i}{2} \triangle, \\ B & = & -i g(t)|U|^2-V_a.
\end{eqnarray}
\end{subequations}
To proceed with split-step type methods it is necessary to compute
the explicit form of the operators $e^{(t-t_0)A}$ and
$e^{(t-t_0)B}$. To get the action of the operators we solve the
subproblems
\begin{subequations}
\label{subpro}
\begin{eqnarray}
\partial_t U & = & \frac{i}{2} \triangle U, \label{A2} \\
\partial_t U & = & - i g(t)|U|^2 U - V_a U. \label{B2}
\end{eqnarray}
\end{subequations}
Since $U \in H^1(\mathbb{R}^2)$, Eq. (\ref{A2}) can be solved in
 Fourier space. Denoting the spatial Fourier transform of $U$
 by
 $\hat{U}(k,t) = \mathcal{F}(U(x,t))$ we get
\begin{equation}
\partial_t \hat{U} = -\frac{i}{2} k^2 \hat{U}, \label{Ak2}
\end{equation}
whose explicit solution is $\hat{U}(k,t) = \hat{U}(k,t_0) e^{-i
k^2 (t-t_0)/2}$.

To solve (\ref{B2}) we write  $U=\rho^{1/2} e^{i\phi}$ and
substituting in Eq. \eqref{B2} we get, from the real part,
$\dot{\rho}=-2V_a\rho$ whose solution is
\begin{equation}\label{rho}
\rho(t)=e^{-2V_a(t-t_0)}\rho(t_0).
\end{equation}
The imaginary part of Eq. \eqref{B2} is $\dot{\phi}=-g(t)\rho$
whose solution is $\phi(t)=\phi(t_0)-\rho(t_0)\int_{t_0}^t g(s)
e^{-2V_a(s-t_0)}ds.$ Finally, we can write
\begin{equation}
U(t)=U(t_0)\exp\left(-V_a(t-t_0)-i\rho(t_0)\int_{t_0}^t g(s)e^{-2
V_a(s-t_0)}ds\right).
\end{equation}
In this way, by defining the time step $\tau$ as $\tau=(t-t_0)/c$,
$c\in\mathbb{R}$, after a suitable renaming we obtain the explicit
form of the operators
\begin{subequations}
\begin{eqnarray}
e^{(t-t_0)A}\equiv e^{c\tau A} & = & \mathcal{F}^{-1} \exp(-i c\tau k^2/2) \mathcal{F},\\
e^{(t-t_0) B}\equiv e^{c\tau B} & = & \exp{\left(-V_a
c\tau-i\rho(t)\int_t^{t+c\tau} g(s) e^{-2 V_a(s-t)}
ds\right)}.\label{expB}
\end{eqnarray}
\end{subequations}
Specifically, when $g(t)$ is given by $g(t)=g_0+g_1 \cos(\Omega
t)$, as we have used in our simulations it is possible to write
explicitly the integral in \eqref{expB} obtaining
\begin{multline}
-\lefteqn{\int_t^{t+c\tau} g(s) e^{-2 V_a(s-t)}ds  =  \frac{g_0}{2
V_a}\left(e^{-2 V_a c\tau}-1\right) -\frac{g_1}{\Omega^2+(2V_a)^2}
\ \times}
 \nonumber \\ \times \left\{e^{-2V_a c\tau}
\left[\Omega\sin\Omega(t+c\tau)
-2V_a\cos\Omega(t+c\tau)\right]-(\Omega\sin\Omega t-2V_a\cos\Omega
 t)\right\}.
\end{multline}
Having the explicit form of the solutions of the subproblems
\eqref{subpro} allows us to obtain the solution of Eq.
\eqref{eq:gpeabsorbe} to any degree of accuracy. We have used two
different splittings: the classical method of second order and a
fourth order optimized method \cite{Blanes} whose equations are
\begin{subequations}
\label{splittings}
\begin{eqnarray}\label{methodnum}
U(x,t+\tau) & = & e^{\tau A/2}e^{\tau B}e^{ \tau A/2} U(
x,t) + \mathcal{O}(\tau^3), \\
U(x,t+\tau) & = & e^{a_1\tau A}e^{b_1\tau B}e^{a_2\tau
A}e^{b_2\tau B}e^{a_3\tau A}e^{b_3\tau B}e^{a_4\tau A}e^{b_3\tau
B}e^{a_3\tau A} \times \nonumber \\& & \times\ e^{b_2\tau
B}e^{a_2\tau A}e^{b_1\tau B}e^{a_1\tau A}U(x,t) +
\mathcal{O}(\tau^5). \label{method4}
\end{eqnarray}
\end{subequations}
with $a_1  = 0.0829844064174052$, $b_1 = 0.245298957184271$, $a_2
= 0.396309801498368$, $b_2 = 0.604872665711080$, $a_3 =
-0.0390563049223486$, $b_3 = 1/2 - (b_1 + b_2)$, $a_4 = 1 - 2(a_1
+ a_2 + a_3)$. Both methods are symmetric compositions of
operators. However, method \eqref{methodnum}
 requires only one
evaluation of $e^A$ and $e^B$ (instead of two) per step because it
is possible to concatenate terms using the First Same As Last
(FSAL) property. For method \eqref{method4} the computational cost
per step is much higher since even using the FSAL property the
number of evaluations of $e^A$ and $e^B$ is six.
\begin{figure}
\begin{center}
\epsfig{file=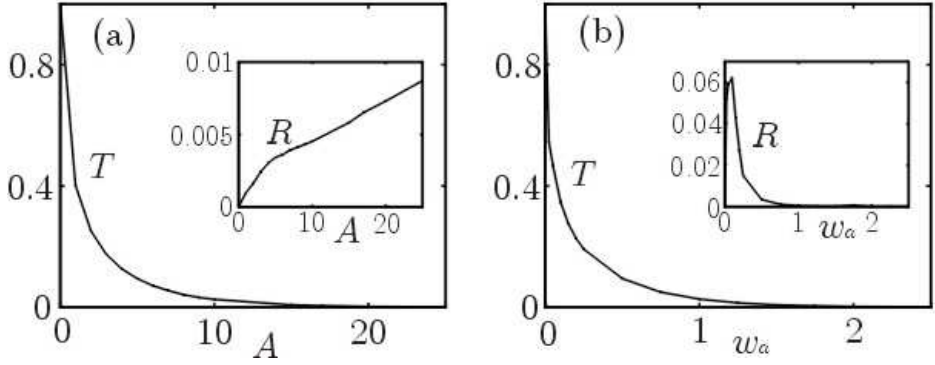,width=0.95\columnwidth}
\end{center}\caption{Reflection ($R$) and transmission ($T$)
coefficients (see text) for the collision of a Townes soliton
 with an absorbing potential $V_a=Ae^{-(r-r_0)^2/2w_a^2}$.
 (a) $T$ and $R$ as a function
of
 $A$  for $w_a=0.5$. (b)
$T$ and $R$ as a function of  $w_a$ for $A=5$.\label{potencial}}
\end{figure}

 These schemes
have many advantages. From the practical point of view the
calculation of the Fourier transform, which is the most computer
time-consuming step in the calculations, may be done by using the
fast Fourier transform (FFT). Thus the computational cost of the
method is of order $\mathcal{O}\left(\mathcal{N}^2\log
\mathcal{N}\right)$, being $\mathcal{N}$ the points number in each
spatial direction of the grid which is quite acceptable. The use
of discrete transforms to represent the continuous Fourier
transform in \eqref{splittings} implicitly imposes periodic
boundary conditions on $U$. However, since $U$ is expected to be
negligible on the boundaries (otherwise the computational domain
must be enlarged or the absorbing potential corrected) this is not
an essential point. Another convenient property of these schemes
is their preservation of the $L^2$-norm of the solutions.

\subsection{The absorbing potential}

 The implementation of transparent boundary conditions is
essential
 to avoid misleading results as it will be clear later. This is because
 of the decomposition of initial data into trapped states plus
some outgoing radiation, a situation which is typical of NLS
problems with fixed coefficients \cite{radiation,radiation2} and
also happens in the case at hand.

Specifically, we have chosen as absorbing potential
$V_a=Ae^{-(r-r_0)^2/2w_a^2}$, where $A, r_0$ and $w_a$
characterize the absorber. The choice of the parameter values must
be done to maximize the absorption of the outgoing radiation. To
choose optimal values we have made several tests by making a
Townes soliton collides with absorbing potentials of several
amplitudes and widths. We have computed the reflection (R) and
transmission (T) coefficients defined as $R=|u_b|/|u_0|$,
$T=|u_f|/|u_0|$ where $u_0$, $u_b$ and $u_f$ are the initial data,
the part of the initial data which moves backwards after
reflection, and the part of the initial data which moves forward
after crossing the absorbing potential, respectively.

Typical examples of the results are shown in Fig. \ref{potencial}.
One must choose a value for $A$ large enough to ensure a good
absorption, whereas the width $w_a$ of the absorbing potential
cannot be chosen too large, since then either the effect on
stabilized structures could be non-negligible or the integration
region should be enlarged to non-practical sizes.

\begin{figure}
\epsfig{file=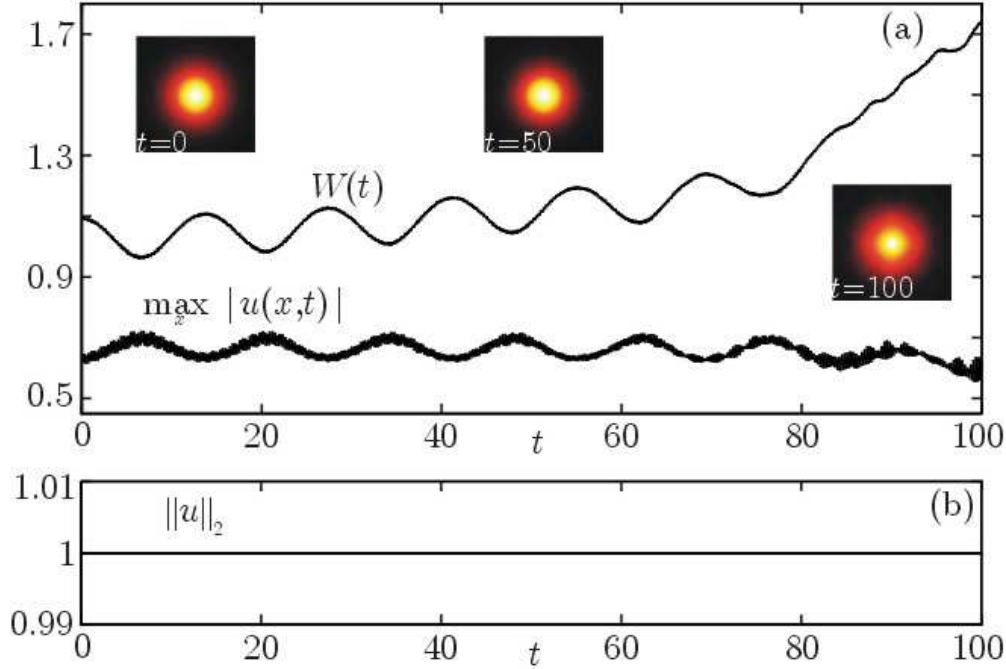,width=0.95\columnwidth} \caption{Results
of numerical simulations of Eq. \eqref{eq:gpe3} showing evolution
of the initial data $u(x,0)=R_{\mu,N}(x)$ with $\mu=0.5$ and
$g=-0.5$ ($W(0)=1.09$) for parameter values $g_0=-2\pi$,
$g_1=8\pi$, $\Omega=40$, \emph{without absorbing potential}. (a)
Evolution of the width $W(t)$ and of the amplitude $\max_x
|u(x,t)|$. The insets show pseudo-color plots of $|u(x,t)|^2$ for
different times. (b) Evolution of the norm
$\|u\|_2$.\label{sinabs}}
\end{figure}

\begin{figure}
\begin{center}
\epsfig{file=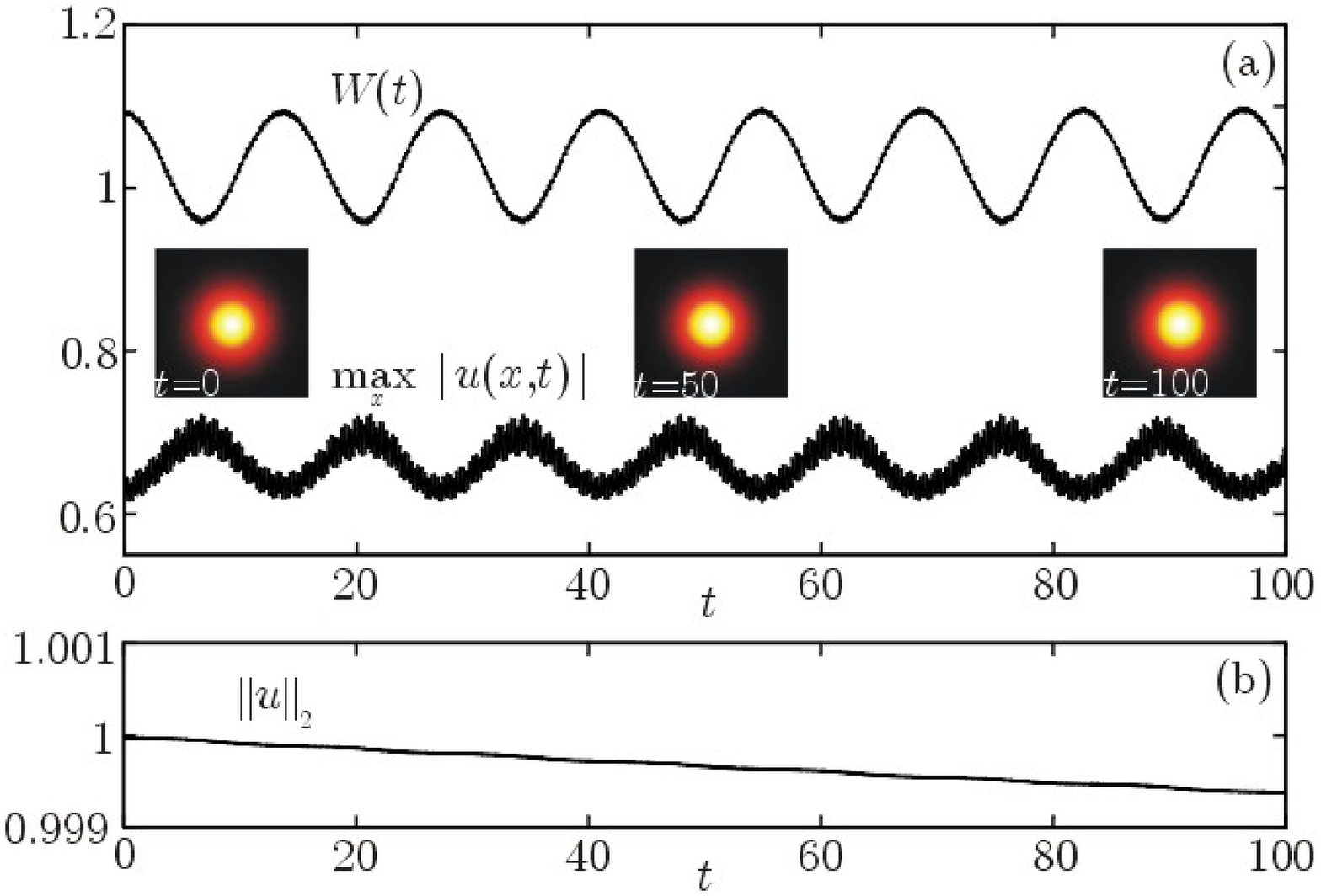,width=0.95\columnwidth}
\end{center}\caption{Same as Fig. \ref{sinabs} but with absorbing potential.
\label{soliton2d}}
\end{figure}

When the absorbing potential is absent the numerical simulations
are misleading and the trapping effects are altered, because the
radiation interacts with the stabilized structure leading to
spurious destabilization. In Figs. \ref{sinabs} and
\ref{soliton2d} we illustrate this behavior starting with a Townes
soliton as initial data.

\section{Stabilization of solitons}\label{stabilization}

\subsection{What is the shape of stabilized solitons?}

\begin{figure}
\begin{center}
\epsfig{file=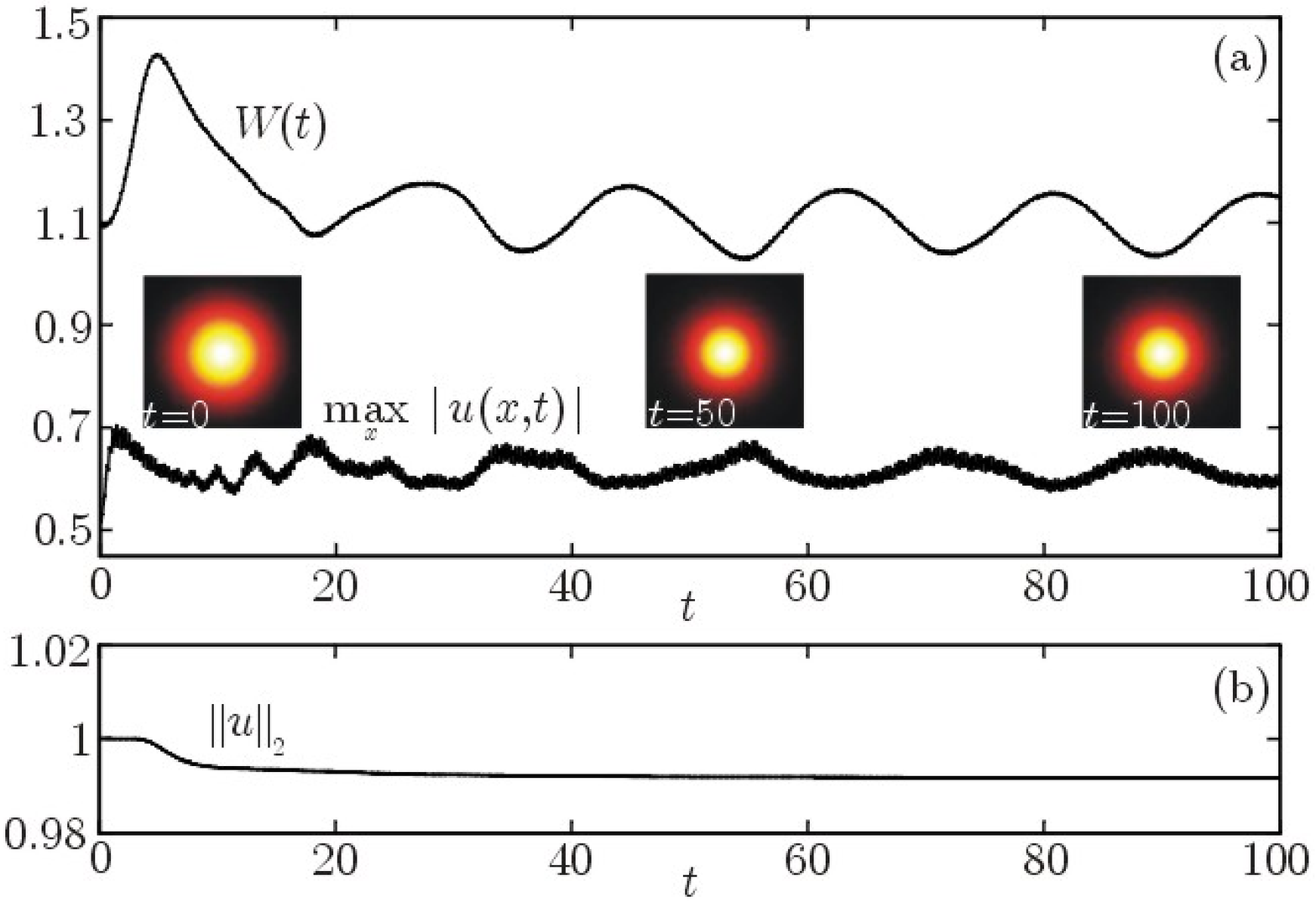,width=0.95\columnwidth}
\end{center}\caption{Results of numerical simulations of Eq. \eqref{eq:gpe3}
showing stabilization of Gaussian initial data
$u(x,0)=(1/\sqrt{\pi}w) e^{-r^2/2 w^2}$ with $w=1.09$ for
parameter values $g_0=-2\pi$, $g_1=8\pi$, $\Omega=40$. (a)
Evolution of the width $W(t)$ and of the amplitude $\max_x
|u(x,t)|$. The insets show pseudocolor plots of $|u(x,t)|^2$ for
different times. (b) Evolution of the norm
$\|u\|_2$.\label{gauss2d}}
\end{figure}

In Refs. \cite{Berge,Malomed,Malomed2,pisaUeda,pisaBoris,Gaspar}
it has been shown that if $g$ is an adequate periodic function
$g(t)$ it is possible to obtain a stabilized structure starting
from different types of initial data, such as Townes solitons or
Gaussian functions. In Figs. \ref{soliton2d} and \ref{gauss2d} we
summarize results taking as initial data a Townes soliton and a
Gaussian function, respectively.

In both cases we get stabilization but in the case of Gaussian
initial data, a readjustment is produced by ejecting outside the
central region a significant part of the wave packet. This happens
during the first 10 time units where the width increases
significantly due to the contribution of the outgoing wave. This
wave is dissipated when hits the absorbing region leading to the
step in the norm evolution shown in Fig. \ref{gauss2d}(b). It was
also conjectured in Ref. \cite{Gaspar} that the stabilized
solution after this process might be a scaled Townes soliton
(which would explain why the stabilization of the Townes soliton
looks smoother). In this section we support this conjecture
quantitatively with precise numerical simulations of Eq.
\eqref{eq:gpe3} using the numerical schemes of Sec. \ref{method}.

To study the shape of the stabilized soliton we define the error
functional
\begin{equation}\label{funcional}
E_{\mu}(t)=\frac{\||u(x,t)|-|R_{\mu}(x)|\|_2}{\|u(x,t)\|_2},
\end{equation}
and the error function $E(t)$ as $E(t) = \min_{\mu}E_{\mu}(t)$.
Note that according to Eq. \eqref{scaling} the parameter $\mu$
parametrizes the shape of the Townes solitons in such a way that
if $\mu$ increases the width of the soliton decreases and the
amplitude increases. The error function $E(t)$ measures the
distance between the solution of Eq. \eqref{eq:gpe3} and the
family of Townes solitons. We have computed numerically the
function $E(t)$ by evaluating the integrals in \eqref{funcional}
for a large set of $\mu$ values and then choosing the minimum
between all of them.

\begin{figure}
\begin{center}
\epsfig{file=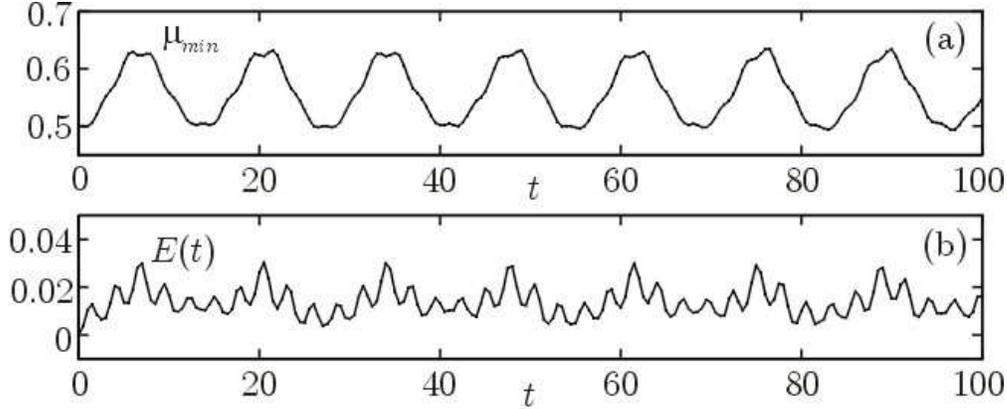,width=0.95\columnwidth}
\end{center}\caption{Analysis of the solutions of Eq. \eqref{eq:gpe3}
for a Townes-soliton initial data. (a) Values of $\mu$ which
minimize the error functional $E_{\mu}$ as a function of time. (b)
Error function $E$ as a function of time.\label{errortownes}}
\end{figure}

In Fig. \ref{errortownes} we summarize our results when a Townes
soliton is taken as initial data. We see, in Fig.
\ref{errortownes}(b)  that $E$ oscillates about a value of $0.02$,
 with a maximum of about $0.03$. This means that the
solution of Eq. (\ref{eq:gpe3}) can be approximated by a Townes
soliton at all times with a maximum error in the norm of about
$3\%$ which is definitely small. In fact, this number could be
even compatible with zero taking into account the number of
approximations involved in our computation: the discretization of
Eq. (\ref{eq:gpe3}), the absorbing potential, the calculation of
the integral \eqref{funcional} and the discrete set of $\mu$
values chosen to compute the minimum.

\begin{figure}
\begin{center}
\epsfig{file=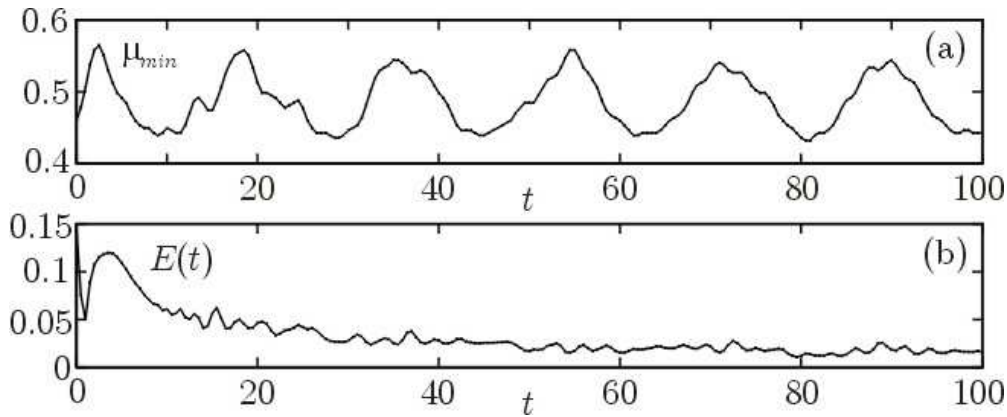,width=0.95\columnwidth}
\end{center}\caption{Same as Fig. \ref{errortownes} but for Gaussian initial data. \label{errorgauss}}
\end{figure}

When Gaussian initial data are taken (Fig. \ref{errorgauss}) we
start with an optimal fitting whose error is about $14\%$. This
error decreases with time to a value around of $2\%$ similar to
the one in the Townes soliton case. This provides a quantitative
support to the idea that Gaussian (or other) initial profiles
evolve into a stabilized Townes soliton by ejecting a fraction of
their amplitude in the form of radiation.

Moreover, in Figs. \ref{errortownes}(a) and \ref{errorgauss}(a) we
see that the minimum points $\mu_{min}$ evolve according to what
we expected. They oscillate in phase opposition respect to the
width evolution of the solutions (see Figs. \ref{soliton2d} and
\ref{gauss2d}), because an increase of the width corresponds to a
decrease of $\mu$ and viceversa (see Eq. \eqref{scaling}).

\subsection{Robustness of stabilized structures to parameter
variations}

\begin{figure}
\begin{center}
\epsfig{file=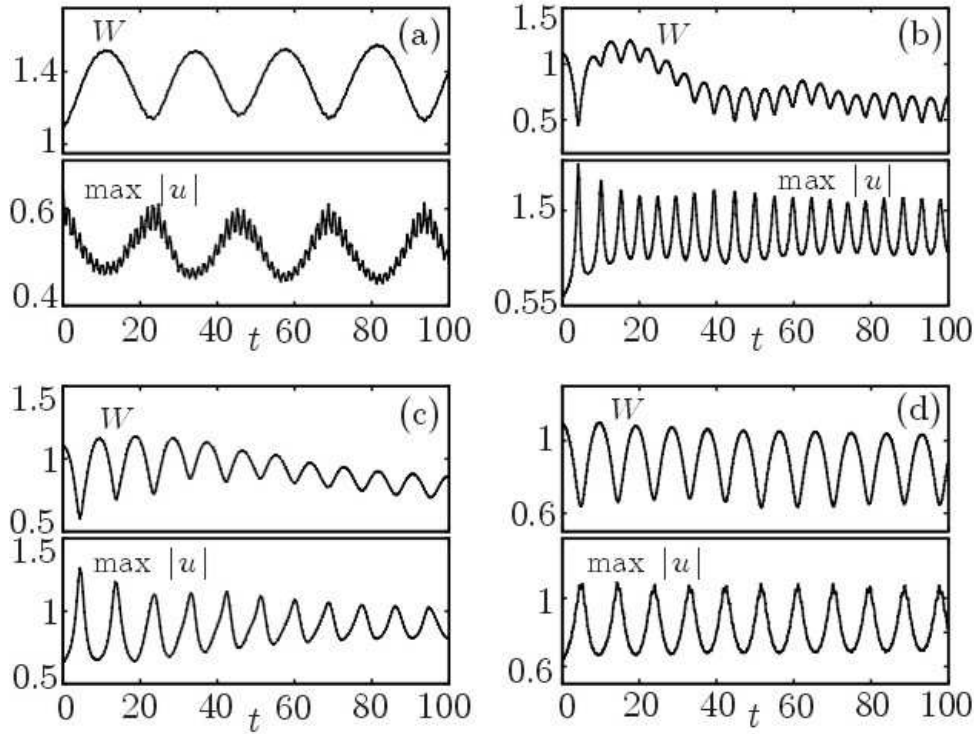,width=0.95\columnwidth}
\end{center}\caption{Stabilization of $u(x,0)=R_{\mu,N}(x)$
with $\mu=0.5$ and $g=-0.5$ ($W(0)=1.09$). Shown are the evolution
of the width $W(t)$ and of the amplitude $\max_x |u(x,t)|$ for
several parameter values. (a) $g_0=-2\pi$, $g_1=6\pi$,
$\Omega=20$. (b) $g_0=-2\pi$, $g_1=8\pi$, $\Omega=250$. (c)
$g_0=-2\pi$, $g_1=6\pi$, $\Omega=80$. (d) $g_0=-2\pi$,
$g_1=15\pi$, $\Omega=150$.\label{barrido}}
\end{figure}

 Finally, to have an idea of the range of parameter
values for which stabilization is possible, we have made
simulations when a Townes soliton is taken as initial data moving
the $g_1$ and $\Omega$ parameters ($g_0=-2\pi$ fixed). We have
found that for $\Omega\in [20,250]$ and $g_1\in [5\pi,15\pi]$ the
stabilization is possible although evolution of the width and of
the amplitude shows different behaviors. In Fig. \ref{barrido} we
plot the results for several parameter values.

\section{Conclusions}

\label{Conclusions}

In this paper we have developed an accurate numerical scheme to
study the stabilization of solutions of the two-dimensional cubic
nonlinear Schr\"{o}dinger equation under variations of the
nonlinear coefficient. We have also given numerical evidences in
favor of the conjecture raised in Ref. \cite{Gaspar} concerning
the shape of stabilized solitons. Finally, we have shown that
stabilization is possible in a wide range of parameter values
which is an interesting result for the possible applications of
these structures to the fields of Bose-Einstein condensation and
nonlinear Optics.

\begin{ack}

This work has been supported by grants BFM2003-02832 (Ministerio
de Ciencia y Tecnolog\'{\i}a) and PAC-02-002 (Consejer\'{\i}a de
Ciencia y Tecnolog\'{\i}a de la Junta de Comunidades de
Castilla-La Mancha). G. D. M. is supported by Ministerio de
Educaci\'on, Cultura y Deporte under grant AP2001-0535.

\end{ack}


\begin{thebibliography}{99}

\bibitem{Sulem}{C. Sulem and P. Sulem, The nonlinear Schr\"{o}dinger equation:
Self-focusing and wave collapse, Springer, Berlin (2000).}

\bibitem{VVz}{L. V\'azquez, L. Streit, V. M. P\'erez-Garc\'{\i}a,
Eds., Nonlinear Klein-Gordon and Schr\"odinger systems: Theory and
Applications, World Scientific, Singapore (1996).}

\bibitem{Fibich}{G. Fibich and G. Papanicolaou, Self-focusing in the perturbed and
unperturbed nonlinear Schr\"{o}dinger equation in critical
dimension, SIAM J. Appl. Math. \textbf{60}, 183-240 (1999).}

\bibitem{Weinstein}{M. I. Weinstein, Non-linear Scr\"{o}dinger-equations and sharp interpolation
estimates, Commun. Math. Phys. \textbf{87}, 567-576 (1983).}

\bibitem{Berge}{L. Berge, V. K. Mezentsev, J. J. Rasmussen, P. L.
Christiansen and Y. B. Gaididei, Self-guiding light in layered
nonlinear media, Opt. Lett. \textbf{25}, 1037-1039 (2000).}

\bibitem{Malomed}{I. Towers and B. A. Malomed, Stable (2+1)-dimensional solitons in a
layered medium with sign-alternating Kerr nonlinearity, J. Opt.
Soc. Am. B {\bf 19}, 537-543 (2002).}

\bibitem{Malomed2}{A. Kaplan, B. V. Gisin, B. A. Malomed, Stable propagation and all-optical
switching in planar waveguide-antiwaveguide periodic structures,
J. Opt. Soc. Am. B \textbf{19}, 522-528 (2002).}

\bibitem{pisaUeda}{H. Saito and M. Ueda, Dynamically stabilized bright
solitons in a two-dimensional Bose-Einstein
condensate, Phys. Rev. Lett. \textbf{90}, 040403 (2003).}

\bibitem{pisaBoris}{F. Abdullaev, J. G. Caputo, R. A. Kraenkel, and B. A.
Malomed, Controlling collapse in Bose-Einstein condensates by
temporal modulation of the scattering length, Phys. Rev. A
\textbf{67}, 013605 (2003).}

\bibitem{Gaspar}{G. D. Montesinos, V. M. P\'erez-Garc\'{\i}a, P.
Torres, Stabilization of solitons of the multidimensional
nonlinear Schr\"{o}dinger equation: Matter-wave breathers, e-print
\texttt{arxiv.org/nlin.PS/0305030} (to appear in Physica D).}

\bibitem{Humberto}{G. D. Montesinos, V. M. P\'erez-Garc\'{\i}a, H.
Michinel, Stabilized two-dimensional vector solitons, e-print
\texttt{arxiv.org/nlin.PS/0310020}.}

\bibitem{Tran}{P. Tran, Solving the time-dependent Schrodinger equation:
Suppression of reflection from the grid boundary with a filtered
split-operator approach, Phys. Rev. E {\bf 58}, 8049-8051 (1998).}

\bibitem{Fevens}{T. Fevens, H. Jiang, Absorbing boundary conditions for the Schr\"{o}dinger equation,
SIAM J. Sci. Comput. {\bf 21}, 255-282
(1999).}

\bibitem{Taha}{T. R. Taha and M. J. Ablowitz, Analytical and numerical aspects of certain
 nonlinear evolution equations .2. Numerical, Nonlinear Schr\"{o}dinger
 equation, J. Comp. Phys. \textbf{55} (1984) 203-230.}

\bibitem{Xiaoyan}{V. M. P\'erez-Garc\'{\i}a and X. Liu, Numerical methods for the simulation of
trapped nonlinear Schrodinger systems, Appl. Math. Comput. {\bf
144}, 215-235 (2003).}

\bibitem{Blanes}{S. Blanes and P. C. Moan, Practical symplectic partitioned Runge-Kutta
and Runge-Kutta-Nystrom methods, J. Comp. Appl. Math. {\bf 142},
313-330 (2002).}


\bibitem{radiation}{S. Cuccagna, Stabilization of Solutions to
Nonlinear Schr\"{o}dinger Equations, Commun. Pure Appl. Math.
\textbf{54}, 1110-1145 (2001).}

\bibitem{radiation2}{T. P. Tsai, H. T. Yau, Asymptotic dynamics of
nonlinear Schr\"{o}dinger equations: Resonance-dominated and
dispersion-dominated solutions, Comm. Pure Appl. Math.
\textbf{55}, 153-216 (2002).}

\end{thebibliography}
\end{document}